# Investigation of the Thermal Stability of the Carbon Framework of Graphene Oxide

Siegfried Eigler*[a], Stefan Grimm[a], Andreas Hirsch[a]

[a] Dr. S. Eigler, Stefan Grimm, Prof. Dr. Andreas Hirsch
Department of Chemistry and Pharmacy and Institute of Advanced Materials and Processes (ZMP)
Friedrich-Alexander-Universität Erlangen-Nürnberg (FAU) Henkestr. 42, 91054 Erlangen and Dr.-Mack Str. 81, 90762
Fürth, Germany
Fax: +49 (0)911 6507865015
E-mail: siegfried.eigler@fau.de

Abstract: In this study, we use our recently prepared graphene oxide (GO) with an almost intact 0-framework of C-atoms (ai-GO) to probe the thermal stability of the carbon framework for the first time. Ai-GO exhibits few defects only by preventing CO2 formation during synthesis. Ai-GO was thermally treated before chemical reduction and subsequently the resulting defect density in graphene was determined by statistical Raman microscopy. Surprisingly, the carbon framework of ai-GO is stable in thin films up to 100 °C. Furthermore, we find evidence for an increasing quality of ai-GO upon annealing at 50 °C before reduction. The carbon framework of GO prepared according to the popular Hummers' method (GO-c) appears to be less stable and decomposition starts at 50 °C what is qualitatively indicated by CO2-trapping experiments in µm-thin films. Information about the stability of GO is important for storing, processing and applying GO in applications.

**Introduction**

Graphene is a nanomaterial that performs best in a large variety of applications that benefit from the high mobility of charge carriers, the transparency or thermal conductivity amongst others.[1] Graphene consists of hexagonal arranged $sp^2$-carbon atoms forming a 2D crystal. Structural defects within this honeycomb lattice must be avoided to make graphene applicable for electronics.[2] Although, especially for applications, it is eligible to produce graphene in large quantities and to process graphene from solution.[3] Despite several auspicious advances regarding graphene production by chemical vapor deposition and solution processing, graphene production still suffers from high temperature synthesis, small flake size or the lack of delamination efficiency.[4] GO does not have these disadvantages and can be produced in large amounts. Thus, it is manifold used as a water-soluble precursor to graphene.[5] GO exhibits hydroxyl, epoxy or organosulfate groups as major species using synthetic protocols that procure on the procedure described by Charpy and Hummers.[6] However, the hexagonal 0-framework of C-atoms in GO is over- oxidized during synthesis using the popular Hummers' method and therefore this conventional GO (GO-c) suffers from defects.[7] Despite manifold reduction techniques, including thermal annealing, the quality of this graphene could not be sufficiently increased.[7]

The distance between defects of reduced GO-c remained below 1-3 nm for the best quality of flakes.[8] We have introduced a production method for GO with an almost intact 0-framework of C-atoms that involves cold processing during synthesis.[9] The mobility of charge carriers of graphene flakes measured after reduction exceeded 1000 cm$^2$/Vs for the best quality of flakes and reflects the high integrity of the honeycomb lattice on the 10 nm scale.[9] Furthermore, with this type of GO the efficiency of reducing agents could be probed by statistical Raman microscopy.[10] Raman spectroscopy turned out to be an adequate characterization tool to determine the defect density of graphene. This is possible because peaks in Raman spectra of graphene vary in intensity and full-width at half-maximum (r) with the distance of defects between 1-20 nm. Thus, ai-GO can be processed and analyzed after reduction to probe the influence of processing.[8a, 9-10, 11]

The thermal stability of GO is a matter of discussion and highly important for commercializing, storing or the processing of GO. In this context, there are several reports that demonstrate the low thermal stability of GO. Hence, GO was found to be metastable at room-temperature and it was further observed that $CO_2$ evolutes already at temperatures between 50 °C and 120 °C in thin films.[12] Moreover, avoiding high temperatures during synthesis resulted in a highly increased quality of GO with an almost intact carbon framework.[9] These investigations suggest that GO is thermally unstable.

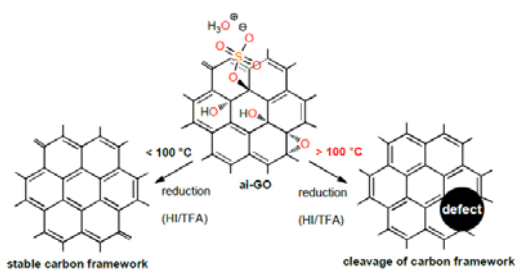

Scheme 1. Schematic representation of the generation of defects in ai-GO due to thermal treatment higher than 100 °C.

Here we find that it is necessary to distinguish between the stability of GO that is related to functional groups and the carbon framework, respectively. Consequently, the carbon framework can be stable at a certain temperature while functional groups are already transformed. That implies that C-C bonds are not cleaved during processing, a prerequisite for GO to be a precursor to graphene. Despite the use of various analyses for GO, involving X-ray photoelectron spectroscopy, X-ray diffraction, infrared spectroscopy, high- resolution transmission electron microscopy and other methods, it was not possible to distinguish between degradation of chemical addends or cleavage of the subjacent carbon framework that obviously produces permanent defects.[7, 13] Here we show that the crucial question of the stability of the carbon framework of GO can indeed be addressed by analyzing reduced ai-GO films by statistical Raman microscopy. We find that films of ai-GO can be thermally treated between room-temperature and 100 °C followed by reduction without causing an additional rupture of the carbon framework as illustrated in **Scheme 1**. At higher temperatures permanent defects are generated that involve missing carbon atoms, or stable oxygen containing structures.[14] Furthermore, we correlate the decomposition temperature of GO determined by Raman spectroscopy with $CO_2$ detection by infrared spectroscopy (FTIR) in films of GO with a thickness of few µm. This $CO_2$-trapping experiment suggests that different types of GO differ in thermal stability with respect to the carbon framework. Moreover, we identify this evolution of $CO_2$ at low temperature as a major source for additional defects in GO.

**Results and Discussion**

At first, ai-GO with an almost intact carbon framework was prepared according to our recently published procedure and ai-GO was coated on Si/300 nm $SiO_2$ substrates by the Langmuir-Blodgett (LB) method.[9] These films of ai-GO were thermally treated for one hour at 20, 30, 40, 50, 75, 100 and 200 °C. After that the GO films were chemically reduced using vapour of hydriodic acid and trifluoro acedic acid at 80 °C.[15] We demonstrated recently that this reduction technique is highly efficient and forms a surface that bears less contaminations compared to other reducing agents, as e.g. ascorbic acid.[10a] In addition, we treated films of GO at 500 °C and 1000 °C in argon. At such high temperatures most of functional groups are removed and an additional reduction step is obsolete since thermal reduction already proceeded and thermally introduced defects can not be healed by chemical reduction.[7, 9]

In **Figure 1** an AFM image of a typical LB-film of graphene flakes with a flake thickness of about 1 nm is shown. The flakes differ in size and shape and overlap partially to form regions of double layers and few layers. For the statistical Raman analysis an area of 100 x 100 µm² was scanned with an increment of 2.5 µm. We recorded more than 1600 Raman spectra and analyzed the peak intensities and the full-widths at half-maximum (r). Raman spectra of reduced GO exhibit three major peaks, the defect induced D peak at about 1335 cm$^{-1}$, the G peak at about 1580 cm$^{-1}$ and the 2D peak at about 2700 cm$^{-1}$. The intensity ratio of the D peak and G peak, as well as the r of the peaks can be used to evaluate the quality of the probed graphene, as introduced by Lucchese et al. and Cançado et al..[11]

While the $I_D/I_G$ ratio follows a relation the r of the D, G and 2D peak continuously increases with the defect density. The relation of the $I_D/I_G$ ratio reaches a maximum that correlates to a distance of defects of about 3 nm and depends on the excitation wavelength used in the Raman experiment. This $I_D/I_G$ maximum is about 4 using a 532 nm laser for excitation of graphene, as we use in this study.[11b] In **Figure S1** mean spectra from an area of 20 x 20 µm² are shown that generally reflect the line broadening with increasing temperature treatment. Furthermore, typical and significant Raman spectra for samples treated at 50 °C, 150 °C, 500 °C and 1000 °C are shown in **Figure S2**.

Values with $r_{2D} < 40$ cm$^{-1}$ are only obtained in the Raman spectra for monolayers of graphene with an average distance of defects larger than about 6 nm. For bi- or few-layers of graphene with the same defect density values of $r_{2D} > 40$ cm$^{-1}$ are measured.[11b, 16] Thus, to remove data from few-layers of graphene it is necessary to analyze predominately monolayers of graphene. This is possible by filtering the dataset according to the intensity of the G peak that is more intense for few-layers of graphene.[9, 11b, 16a] Following this procedure we finally use about 800-1000 Raman spectra for the statistical analysis of each sample. Detailed data of the mean $I_D/I_G$ ratio and the r are shown in **Figure 3**, **Figure S3 and Table S1**. Furthermore, we plotted the $I_D/I_G$ ratio against the $r_{2D}$ in **Figure 2.**

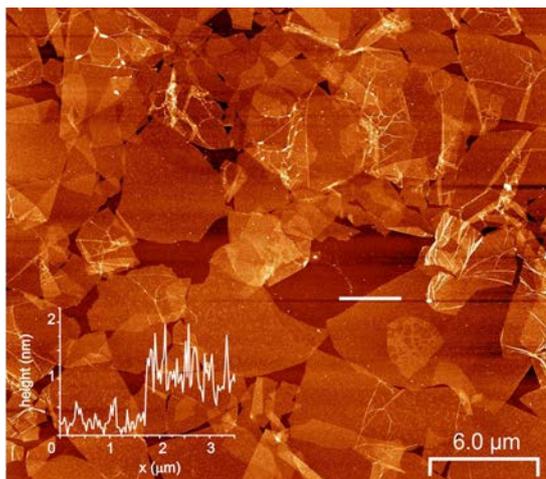

Figure 1. AFM image of a Langmuir-Blodgett film of graphene flakes on $SiO_2$ obtained after reduction of ai-GO; scanned size 30 x 30 µm; average diameter of flakes about 5 µm.

We prepared ten different samples coated with ai-GO that exhibit the potential to be reduced to graphene that can be characterized by a mean $I_D/I_G$ ratio of 2.6±0.3 and a $r_{2D}$ of 74±12 cm$^{-1}$. With storing the GO coated wafers at temperatures higher than 100 °C the quality of reduced ai-GO decreases dramatically. Thus, the 150 °C sample bears additional defects that are due to the loss of carbon what is indicated by thermogravimetric bulk analysis (TGA) of GO.[9, 12a] The $r_{2D}$ of 182±66 cm$^{-1}$ with its broad standard deviation reflects the introduction of defects in the carbon framework. The heterogeneity of this sample is deduced in **Figure 2** (blue dots). While some flakes of ai-GO remained partially intact ($r_{2D} < 75$ cm$^{-1}$) others with a $r_{2D} > 200$ cm$^{-1}$ indicate the degraded hexagonal carbon framework. At 200 °C ai-GO is thermally reduced or more likely decomposes what correlates to the main weight-loss detected by TGA.[9] The high defect density in graphene is reflected by $r_{2D}=225±26$ cm$^{-1}$. However, the highest defect density is determined after annealing ai-GO at 500 °C ($r_{2D}=301±40$ cm$^{-1}$, green dots in **Figure 2**) and partial reconstruction of the graphene lattice can be observed after annealing ai-GO at 1000 °C ($r_{2D}= 127±40$ cm$^{-1}$, purple dots in **Figure 2**). Nevertheless, the integrity of the carbon framework for samples treated :S 100 °C was not reached by thermal reduction. Thus, annealing ai-GO at 1000 °C is not a viable way to graphene (**Figure 3**, **Table S1**).[17]

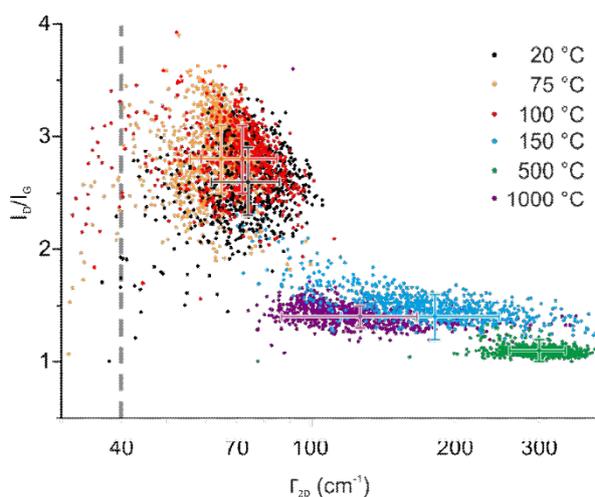

Figure 2. Plot of $I_D/I_G$ against r2D of thermally treated and afterwards chemically reduced ai-GO films (20, 75, 100 and 150 °C) and thermally reduced ai-GO at 500 °C and 1000 °C; error bars are derived from the standard deviation.

An intriguing and surprising result is reflected by the Raman values obtained after thermal annealing between 30 °C and 100 °C. The best quality of flakes is not obtained from the sample stored at 20 °C but for samples treated between 40 °C and 100 °C. Thus, spectra with a $r_{2D} < 40$ cm$^{-1}$, a prerequisite for a high quality of graphene, are predominately obtained for tempered samples, as shown in **Figure 2** (orange dots and red dots, 50 °C and 100 °C, detailed plots in **Figure S4**). To quantify the amount of high-quality spectra we prepared $r_{2D}$ histograms that surprisingly and reproducible reveal an about doubled percentage (from 3% to 7%) of high-quality flakes for films treated at a temperature of 50 °C before reduction (**Figures S5**). Further on, we extract a $r_{2D}$ of 64±9 cm$^{-1}$ (50 °C sample) instead of 74±12 cm$^{-1}$ (20 °C sample). Even if this difference is small we propose that defects are rather chemically healed than formed by treating the sample at 50 °C. This result is surprising and contradictory to the observation that GO-c already decomposes at 50 °C accompanied by $CO_2$ formation.[12a] This $CO_2$ can be trapped in films of GO of few µm thickness during heating, accumulates and forms blisters that burst at 120 °C. Such a loss of carbon denotes the introduction of a defect within the carbon framework and thus further limits the quality of graphene after reduction. To clarify this antagonism we repeated the FTIR study using the ai-GO and find that trapping of $CO_2$ occurs between 90 °C and 160 °C for ai-GO instead of 50 °C and 140 °C as observed for GO-c. This trapping experiment also explains the poor performance of graphene films obtained after 150 °C treatment because $CO_2$ already evolutes from the carbon framework (**Figure 2**). Thus, the $CO_2$-trapping experiment and the statistical Raman microscopic results coincide and the thermal decomposition temperature is indicated by both methods between about 90 °C and 100 °C. Hence, this experimentally detected $CO_2$ acts as a source for defects in graphene. Our investigations implicate that the carbon framework of GO is stable up to about 100 °C. However, this is only valid for ai-GO because the $CO_2$-trapping experiment for GO-c indicates $CO_2$ formation already at 50 °C. Unfortunately, statistical Raman microscopy can not be applied on reduced GO-c because it already exhibits too many defects. Regrettably, the structural motives responsible for the evolution of $CO_2$ from GO at low temperature remain unknown but $CO_2$ is most likely formed within highly oxidized regions of GO. Further research to identify thermally labile functional groups in GO is highly desired to further improve the quality GO.

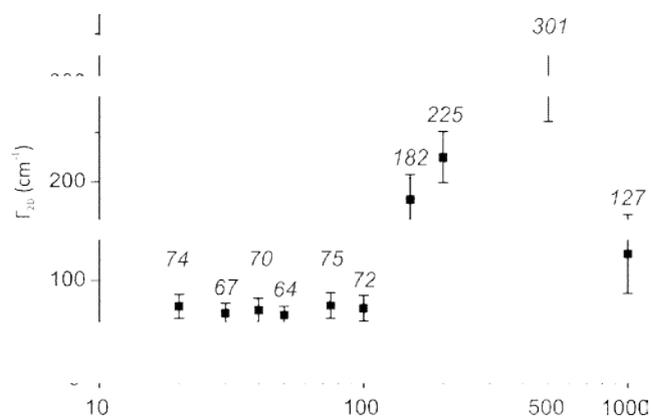

Figure 3. Plot of r2D against pretreatment-temperature of ai-GO films before reduction; italic numbers are mean r2D values, error bars are derived from the standard deviation.

**Conclusions**

According to the data presented here it is evident that the carbon framework of GO is thermally more stable than functional groups that have been thoroughly analyzed in the literature.[12b, 12c, 13a, 18] We conclude that the stability of GO needs to be addressed either to the chemical addends of GO or the hexagonal 0-framework of C-atoms, which we addressed here. We used ai-GO as starting material and analyzed processed and reduced monolayers of ai-GO by statistical Raman microscopy. Surprisingly, we found that the carbon framework of ai-GO is thermally stable up to 100 °C. Furthermore, we deduce on a slightly increased quality of graphene by treating GO at about 50 °C before reduction. In addition, we correlate the thermal stability extracted from statistical Raman microscopy to the $CO_2$-trapping experiment based on FTIR spectroscopy of µm-thin GO films. The results suggest that the carbon framework of GO-c prepared according to Hummers' method is less stable ($CO_2$- formation starts at about 50 °C) than that of ai-GO ($CO_2$-formation starts at about 90 °C). Therefore, this technique may also be useful to qualitatively determine the quality of GO after bulk production. Since GO must be storable and processible under ambient conditions or even at elevated temperatures the thermal stability of the carbon framework of GO is highly important for the development of GO based applications and should be considered in future studies.

**Experimental Section**

**General Methods.** Natural flake graphite was obtained from Asbury Carbon. The grade used was 3061. Potassium permanganate, sodium nitrate, sulfuric acid, hydriodic acid and trifluoroacetic acid were obtained from Sigma-Aldrich®. Freeze-drying was accomplished on an ALPHA 1-4 LDplus from Matrtin Christ, Germany. For centrifugation a Sigma 4K15 centrifuge, Sigma Laborzentrifugen GmbH, Germany, was used. Elemental analysis was performed by combustion and gas chromatographic analysis with a VarioMicro CHNS analyzer from Elementar Analysensysteme GmbH, Hanau, Germany. Langmuir-Blodgett films were prepared using Langmuir-Blodgett Minitrough from KSV NIMA and films are prepared from MeOH/water mixtures on water as subphase at a pressure of 1.5 mN/m to yield slightly over packed films. These films were used for atomic force microscopy (AFM) imaging and scanning Raman spectroscopy. AFM measurements were performed in tapping mode using SolverPro from NT-MDT. Thermogravimetric analysis (TGA) equipped with a mass spectrometer (MS) was accomplished on a Pyris 1 TGA equipped with Clarus SQ 8 C mass spectrometer with the following programmed time dependent temperature profile: RT- 700 °C (TGA-MS) with 10 K/min gradient, and cooling to RT. The initial sample weights were about 1.5 mg and the whole experiment was accomplished under inert gas
atmosphere with a $N_2$ gas flow of 70 ml/min. For thermal annealing up to 1000 °C in argon atmosphere a tube furnace B 180 from Nabertherm, Germany was used, equipped with a quartz tube filled with the sample and argon. Heating rate was 50 °C/min and the samples were kept for 1 h at the desired temperature. Bruker Tensor FTIR spectrometer
equipped with ZnSe was used for the measurements of GO films.

**Preparation of GO (ai-GO).** The method for preparation is adopted from an earlier publication.[9] No pre-treatment of graphite was applied. Graphite (1 g, 83 mmol grade 3061, Asbury Carbon) and sodium nitrate (0.5 g, 5.9 mmol) were dispersed in concentrated sulphuric acid (24 mL). The dispersion was cooled to about 0 °C. After that, over a period of three hours potassium permanganate (3.0 g, 19 mmol) was added. The temperature of the reaction mixture was kept below 10 °C and stirred for an additional 16 h. The reaction mixture was still cooled and during cooling diluted sulphuric acid (20 mL, 10 %) was continuously added to the reaction mixture over 2 h. After that water was added (60 mL, 6 h) and the temperature of the reaction mixture was kept below 10 °C. The reaction mixture was then poured on ice (500 mL) and hydrogen peroxide (20 mL, 3 %) was added drop wise until gas evolution was completed, and the temperature was kept below 10 °C. The obtained graphite oxide was purified by repeated centrifugation and dispersion in cooled water (below 10 °C) until the pH of the supernatant was neutral. Finally, GO was yielded by mild sonication using a bath sonicator. Even without sonication, graphite oxide exfoliated to GO in some extent. The suspension was finally centrifuged three times at 5,000g to remove remaining graphite oxide. Elemental Analysis: C 45.68, H 2.28, N 0.03, S 3.62. Total
weight loss according to thermogravimetric analysis ($N_2$, 25-700 °C): 51.0 %;  yield: 750 mg freeze dried.

**Preparation of conventional GO (GO-c).** The preparation was accomplished according to the standard method described by Hummers and Offeman and the method was described before (including analytical data).[12a]

**GO films on ZnSe.** A diluted dispersion of GO was drop-casted on a ZnSe window and the film was formed by evaporation of the solvent at ambient conditions. The thickness of the film was determined using the Z-indicator of the Zeiss microscope. Thereto, the film was scratched and either the top of the film or the substrate was focused. Typically, films with a thickness of 2 µm were formed.[12a]

**Reduction of Langmuir-Blodgett films on Si/SiO$_2$ wafers.** The wafers were paced in a vial on glass wool and some drops of hydriodic acid (57% in water) were placed on the glass wool. After that some drops of trifluoro acetic acid were dropped on the glass wool also. The vial was closed and after five minutes at room temperature the vial was
heated to 80 °C for an additional 10 minutes. After that the wafers were rinsed with water and dried at 80 °C.

**Acknowledgements**

The authors thank the Deutsche Forschungsgemeinschaft (DFG - SFB 953, Project A1 "Synthetic Carbon Allotropes"), the European Research Council (ERC; grant 246622 - GRAPHENOCHEM), and the Cluster of Excellence 'Engineering of Advanced Materials (EAM)' for financial support.

[1] a) K. S. Novoselov, V. I. Fal'ko, L. Colombo, P. R. Gellert, M. G. Schwab, K. Kim, *Nature* **2012**, *490*, 192-200; b) C. Chung, Y. K. Kim, D. Shin, S. R. Ryoo, B. H. Hong, D. H. Min, *Acc. Chem. Res.* **2013**, *46*, 2211-2224; c) Z. Liu, A. A. Bol, W. Haensch, *Nano Lett.* **2011**, *11*, 523-528; d) E. Pop, V. Varshney, A. K. Roy, *MRS Bull.* **2012**, *37*, 1273-1281; e) H. Y. Mao, S. Laurent, W. Chen, O. Akhavan, M. Imani, A. A. Ashkarran, M. Mahmoudi, *Chem. Rev.* **2013**, *113*, 3407-3424.

[2] M. Wang, S. K. Jang, W. J. Jang, M. Kim, S. Y. Park, S. W. Kim, S. J. Kahng, J. Y. Choi, R. S. Ruoff, Y. J. Song, S. Lee, *Adv. Mater.* **2013**, *25*, 2746-2752.

[3] D. R. Cooper, B. D'Anjou, N. Ghattamaneni, B. Harack, M. Hilke, A. Horth, N. Majlis, M. Massicotte, L. Vandsburger, E. Whiteway, V. Yu, *ISRN Condens. Matter Phys.* **2012**, *2012*, 1-56.

[4] F. Bonaccorso, A. Lombardo, T. Hasan, Z. Sun, L. Colombo, A. C. Ferrari, *Mater. Today* **2012**, *15*, 564-589.

[5] D. Chen, H. Feng, J. Li, *Chem. Rev.* **2012**, *112*, 6027-6053.

[6] a) G. Charpy, *C. R. Hebd. Séances Acad. Sci.* **1909**, *148*, 920-923; b) J. William S. Hummers, R. E. Offeman, *J. Am. Chem. Soc.* **1958**, *80*, 1339; c) S. Eigler, C. Dotzer, F. Hof, W. Bauer, A. Hirsch, *Chem. Eur. J.* **2013**, *19*, 9490-9496.

[7] S. Mao, H. Pu, J. Chen, *RSC Adv.* **2012**, *2*, 2643-2662.

[8] a) S. Eigler, C. Dotzer, A. Hirsch, *Carbon* **2012**, *50*, 3666-3673; b) G. Eda, J. Ball, C. Mattevi, M. Acik, L. Artiglia, G. Granozzi, Y. Chabal, T. D. Anthopoulos, M. Chhowalla, *J. Mater. Chem.* **2011**, *21*, 11217-11223.

[9] S. Eigler, M. Enzelberger-Heim, S. Grimm, P. Hofmann, W. Kroener, A. Geworski, C. Dotzer, M. Rockert, J. Xiao, C. Papp, O. Lytken, H. P. Steinruck, P. Muller, A. Hirsch, *Adv. Mater.* **2013**, *25*, 3583-3587.

[10] a) S. Eigler, S. Grimm, M. Enzelberger-Heim, P. Muller, A. Hirsch, *Chem. Commun.* **2013**, *49*, 7391-7393; b) J. M. Englert, P. Vecera, K. C. Knirsch, R. A. Schafer, F. Hauke, A. Hirsch, *ACS Nano* **2013**, *7*, 5472-5482.

[11] a) M. M. Lucchese, F. Stavale, E. H. M. Ferreira, C. Vilani, M. V. O. Moutinho, R. B. Capaz, C. A. Achete, A. Jorio, *Carbon* **2010**, *48*, 1592-1597; b) L. G. Cançado, A. Jorio, E. H. M. Ferreira, F. Stavale, C. A. Achete, R. B. Capaz, M. V. O. Moutinho, A. Lombardo, T. S. Kulmala, A. C. Ferrari, *Nano Lett.* **2011**, *11*, 3190-3196.

[12] a) S. Eigler, C. Dotzer, A. Hirsch, M. Enzelberger, P. Müller, *Chem. Mater.* **2012**, *24*, 1276-1282; b) S. Kim, S. Zhou, Y. Hu, M. Acik, Y. J. Chabal, C. Berger, W. de Heer, A. Bongiorno, E. Riedo, *Nature Mater.* **2012**, *11*, 544-549; c) M. Acik, C. Mattevi, C. Gong, G. Lee, K. Cho, M. Chhowalla, Y. J. Chabal, *ACS Nano* **2010**, *4*, 5861-5868.

[13] a) S. H. Huh, in *Physics and Applications of Graphene - Experiments* (Ed.: S. Mikhailov), InTech, **2011**, pp. 73-90; b) C. Gómez-Navarro, J. C. Meyer, R. S. Sundaram, A. Chuvilin, S. Kurasch, M. Burghard, K. Kern, U. Kaiser, *Nano Lett.* **2010**, *10*, 1144-1148; c) Y. Yamada, H. Yasuda, K. Murota, M. Nakamura, T. Sodesawa, S. Sato, *J. Mater. Sci.* **2013**, *48*, 8171-8198.

[14] C. Botas, P. Álvarez, P. Blanco, M. Granda, C. Blanco, R. Santamaría, L. J. Romasanta, R. Verdejo, M. A. López-Manchado, R. Menéndez, *Carbon* **2013**, *65*, 156-164.
[15] P. Cui, J. Lee, E. Hwang, H. Lee, *Chem. Commun.* **2011**, *47*, 12370-12372.
[16] a) S. Chen, W. Cai, R. D. Piner, J. W. Suk, Y. Wu, Y. Ren, J. Kang, R. S. Ruoff, *Nano Lett.* **2011**, *11*, 3519-3525; b) S. Lee, K. Lee, Z. Zhong, *Nano Lett.* **2010**, *10*, 4702-4707.
[17] a) D. Yang, A. Velamakanni, G. Bozoklu, S. Park, M. Stoller, R. D. Piner, S. Stankovich, I. Jung, D. A. Field, C. A. V. Jr., R. S. Ruoff, *Carbon* **2009**, *47*, 145-152; b) C. Mattevi, G. Eda, S. Agnoli, S. Miller, K. A. Mkhoyan, O. Celik, D. Mastrogiovanni, G. Granozzi, E. Garfunkel, M. Chhowalla, *Adv. Funct. Mater.* **2009**, *19*, 2577-2583.
[18] S. Eigler, S. Grimm, F. Hof, A. Hirsch, *J. Mater. Chem. A* **2013**, *1*, 11559-11562.

**INVESTIGATION OF THE THERMAL STABILITY OF THE CARBON FRAMEWORK OF GRAPHENE OXIDE**

Siegfried Eigler,*[a] Stefan Grimm,[a] Andreas Hirsch[a]

[a] Department of Chemistry and Pharmacy and Institute of Advanced Materials and Processes (ZMP), Friedrich-Alexander-Universität Erlangen-Nürnberg (FAU), Dr.-Mack Str. 81, 90762 Fürth, Germany. Fax: +49 (0)911 6507865015; Tel: +49 (0)911 6507865005; E-mail: siegfried.eigler@fau.de

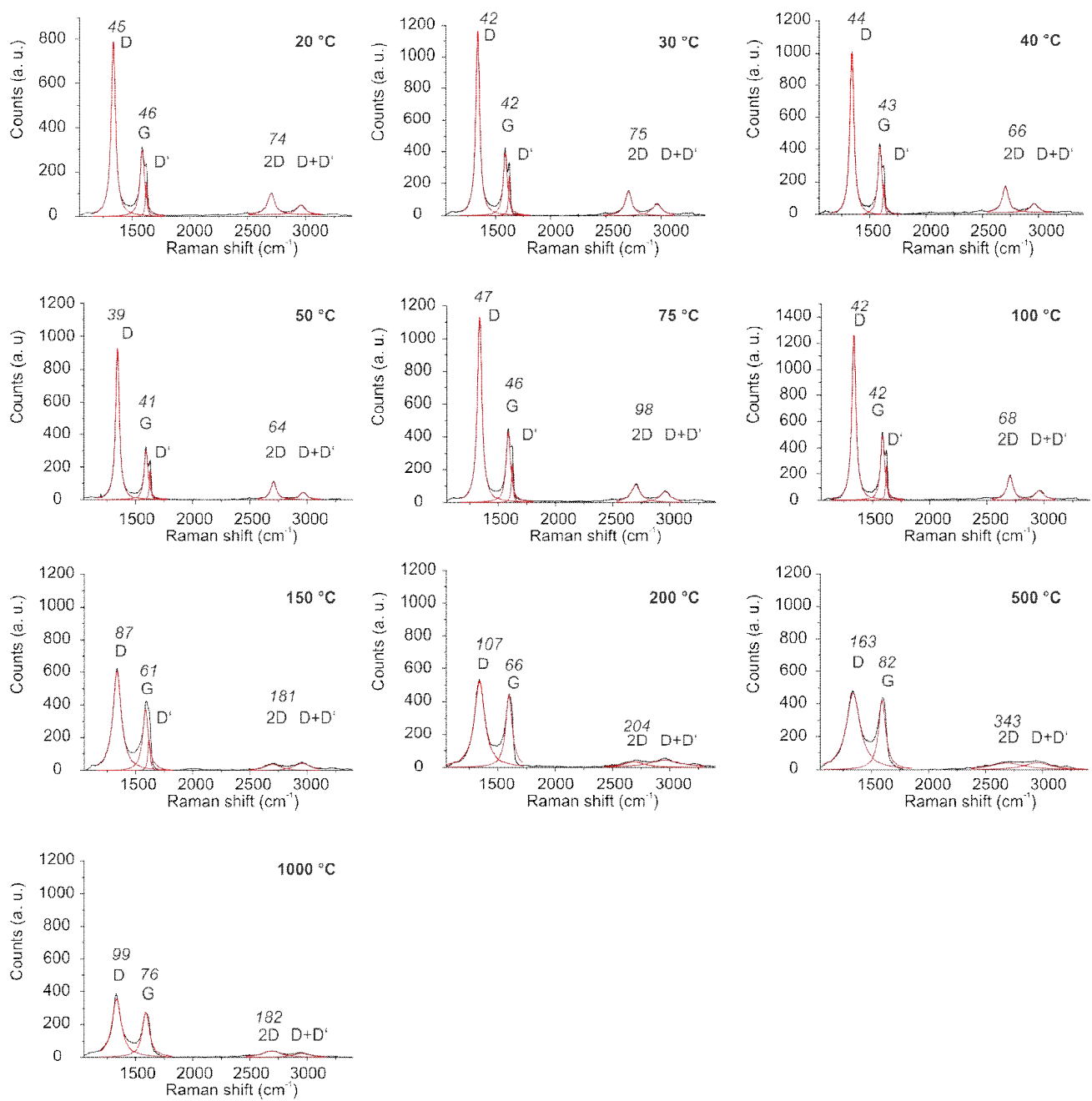

Figure S1. Mean Raman spectra of reduced ai-GO pretreated at different temperatures (20 °C, 30 °C, 40 °C, 50 °C, 75 °C, 100 °C, 200 °C, 500 °C and 1000 °C. The full-width at half-maximum of the D peak, G peak and 2D peak are indicated by the numbers in italic; the lorentzian fit of peaks is shown in red.

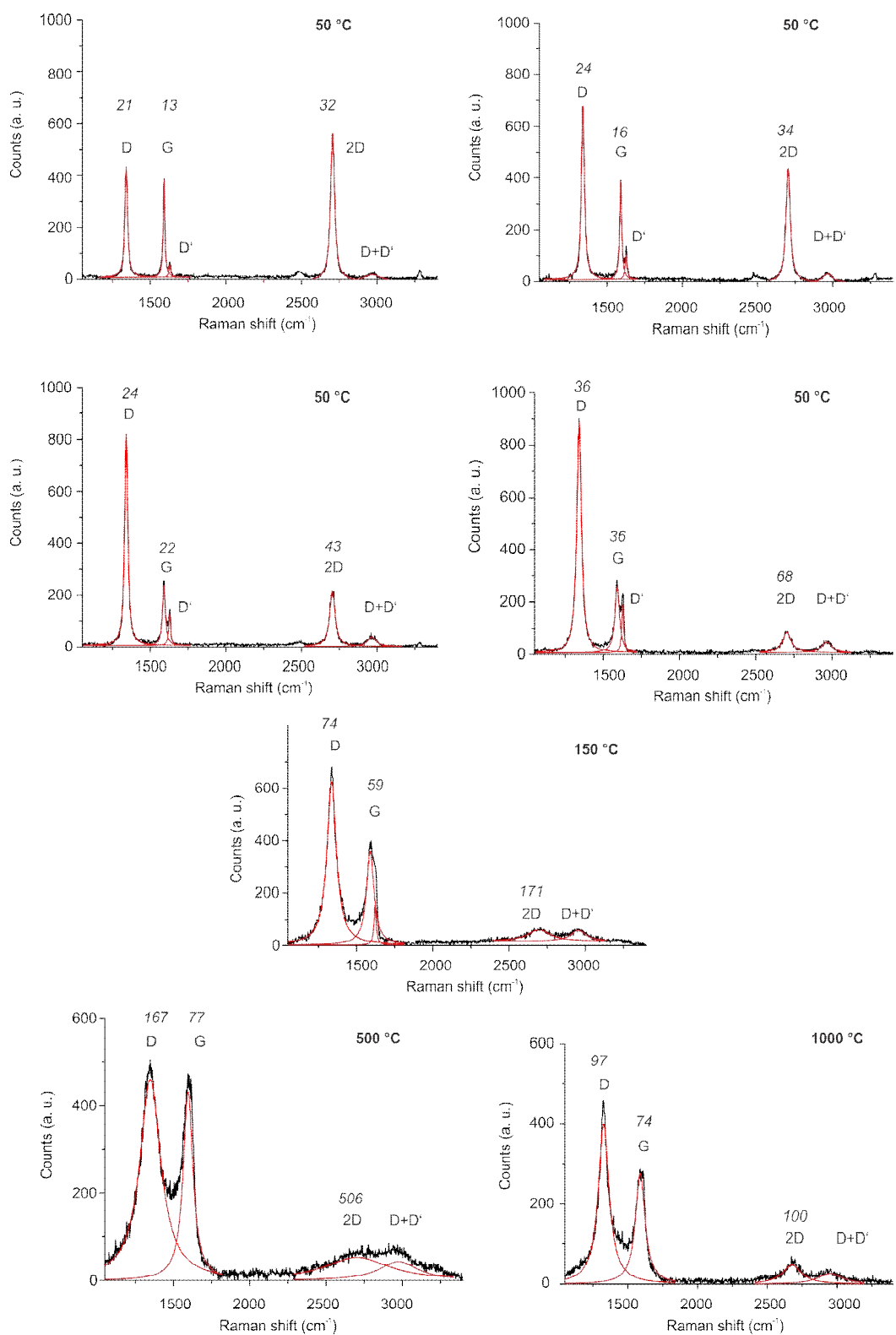

Figure S2. Plots of typical Raman spectra from reduced ai-GO pretreated at different temperatures 50 °C, 150 °C, 500 °C and 1000 °C. The full-width at half-maximum of the D peak, G peak and 2D peak are indicated by the numbers in italic; the lorentzian fit of peaks is shown in red.

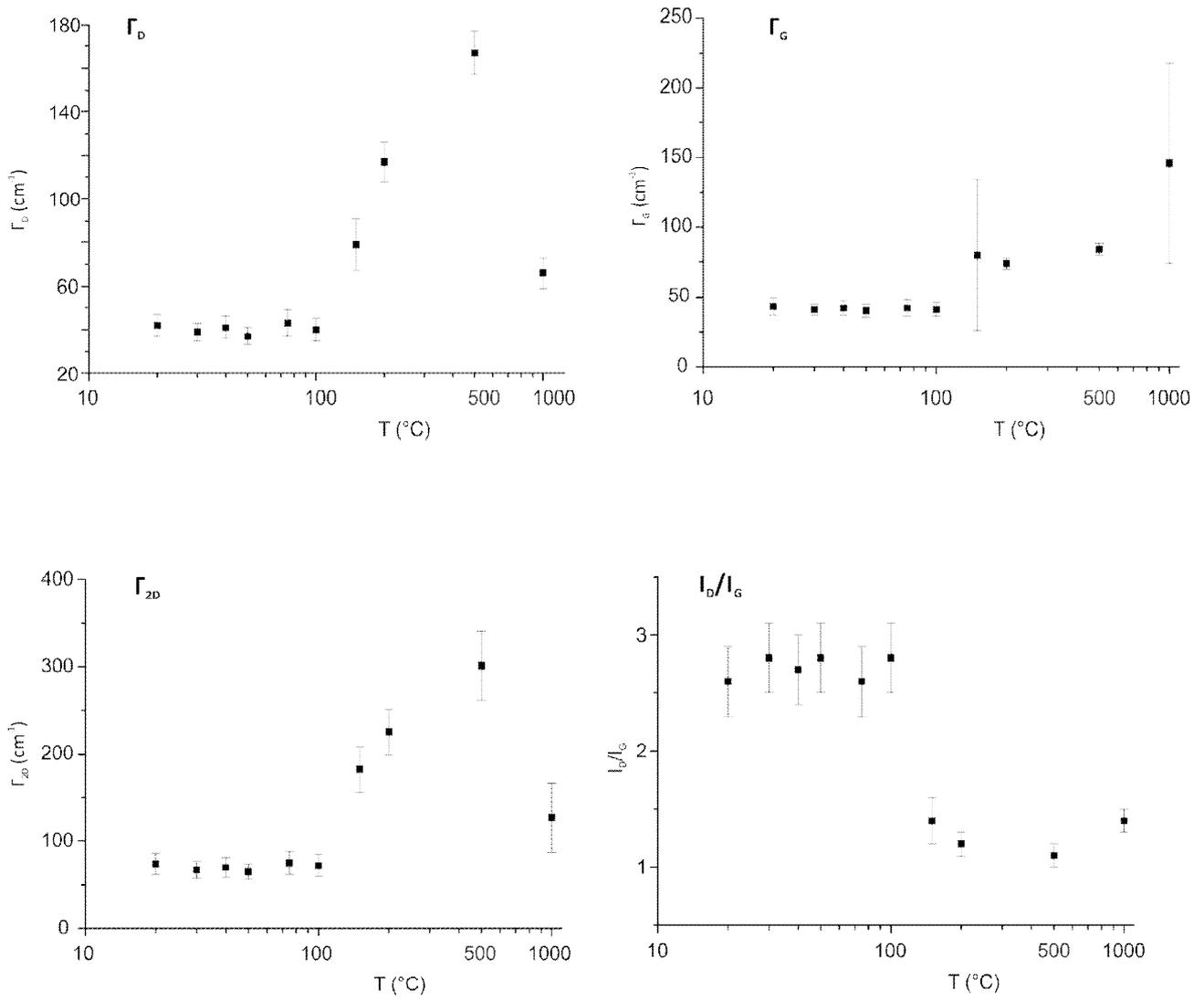

Figure S3. Plot of the full-width at half-maximum (Γ) of the D peak, G peak and the 2D peak and the $I_D/I_G$ ratio of reduced ai-GO pre-treated at different temperatures 20 °C, 30 °C, 40 °C, 50 °C, 75 °C, 100 °C, 150 °C, 200 °C, 500 °C and 1000 °C.

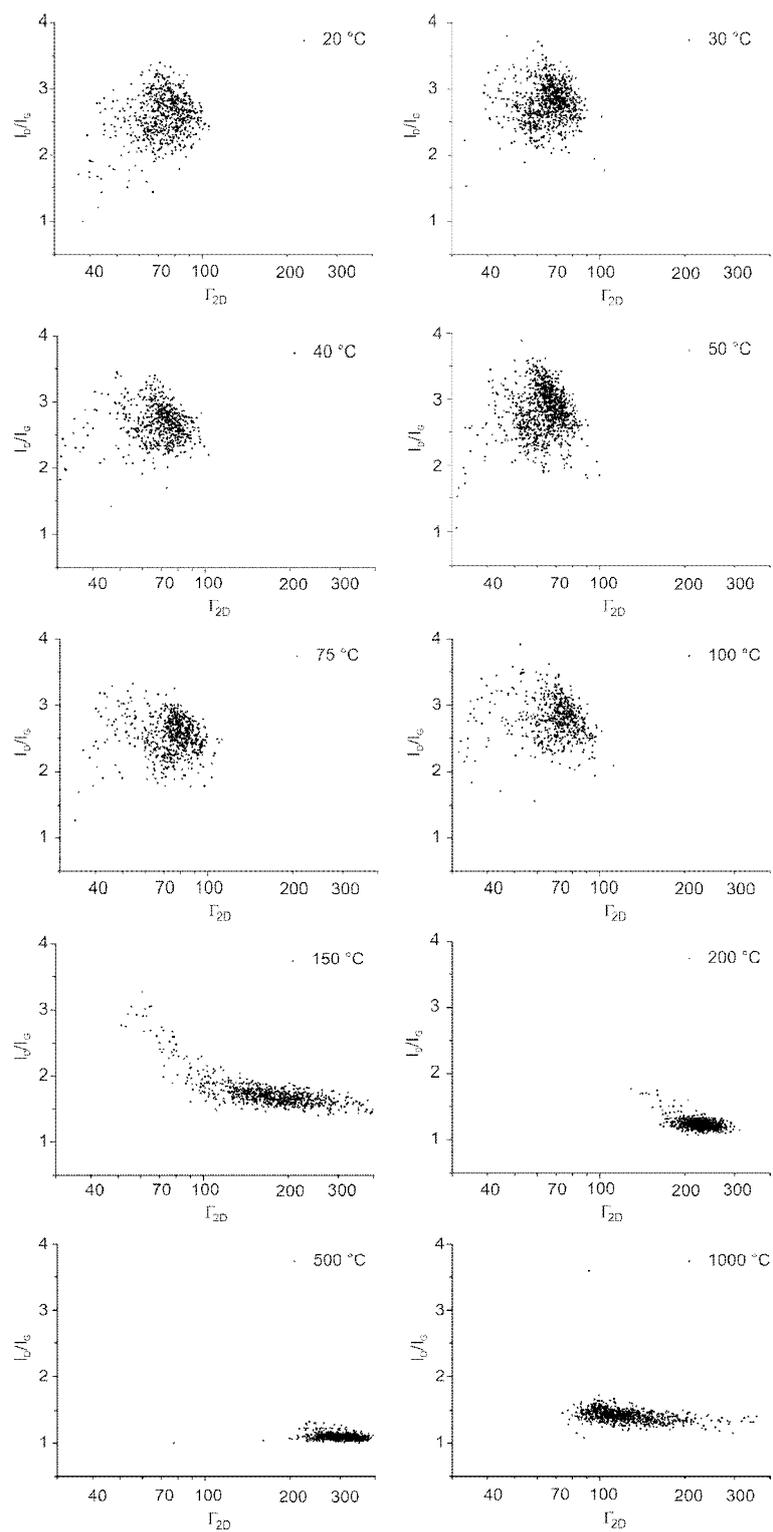

Figure S4. Plot of $I_D/I_G$ ratio against the full-width at half-maximum of 2D peak ($\Gamma_{2D}$) of reduced ai-GO pretreated at different temperatures 20 °C, 30 °C, 40 °C, 50 °C, 75 °C, 100 °C, 150 °C, 200 °C, 500 °C and 1000 °C.

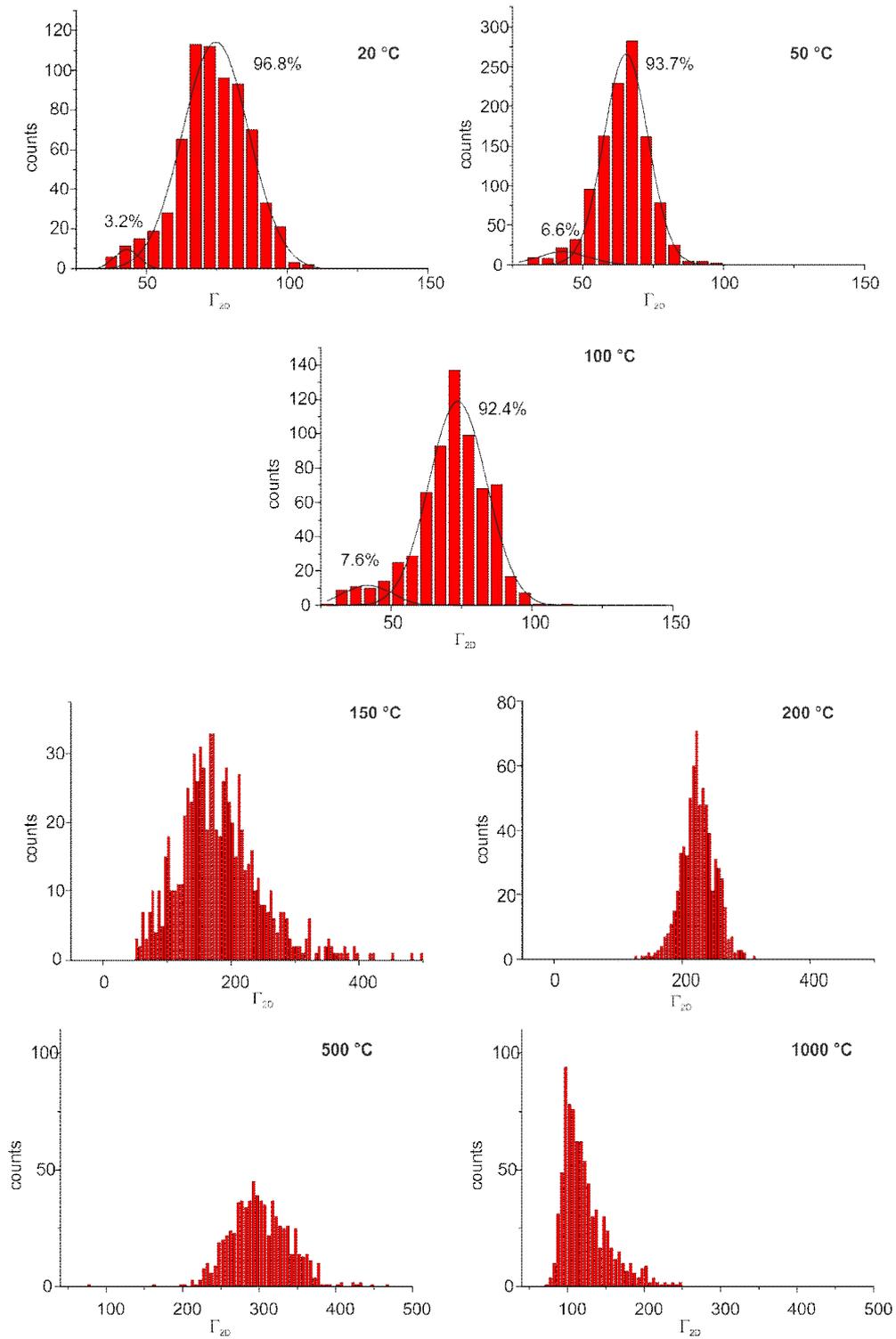

Figure S5. Histograms of the full-width at half-maximum of the 2D peak ($\Gamma_{2D}$) of reduced aiGO pretreated at different temperatures (20 °C, 50 °C, 100 °C, 150 °C, 200 °C, 500 °C and 1000 °C).

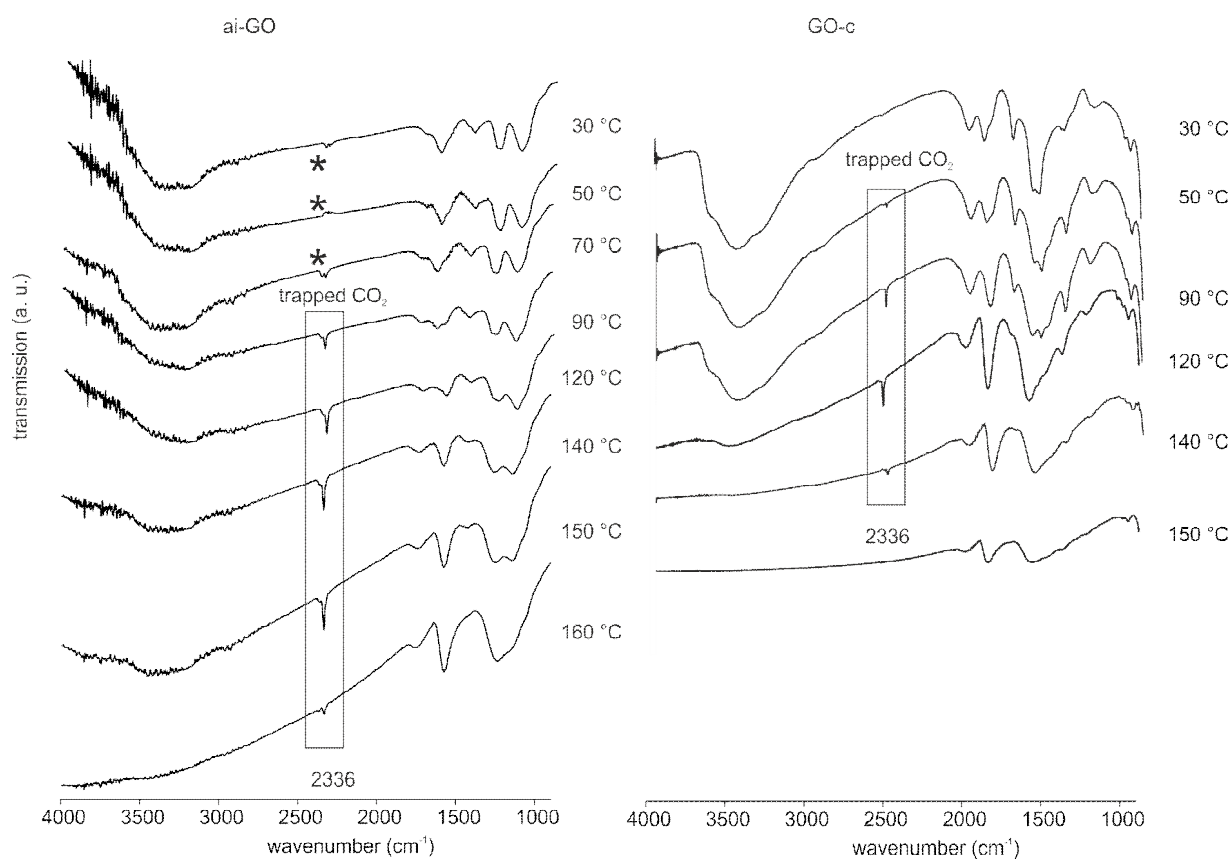

Figure S6. FTIR spectra of GO after thermal treatment between 30 °C and 160 °C; left: ai-GO: $CO_2$ is trapped between layers of GO between about 90 °C and 160 °C, (at 70 °C the shape of $CO_2$ vibration ins symmetrical and was therefore assigned to ambient $CO_2$); right: GO-c: $CO_2$ is trapped between layers of GO between about 50 °C and 140 °C (* vibrations of $CO_2$ with low intensity that originate from ambient $CO_2$, for GO-c at 50 °C the shape of the $CO_2$ signal is unsymmetrical indicating tapped $CO_2$); this trapping experiment is also described in detail in an earlier publication.[1]

Table S1. $I_D/I_G$ ratio and full-width at half-maximum of the D peak, G peak and 2D peak of thermally treated and reduced ai-GO determined by statistical Raman microscopy.

| T (°C) | $I_D/I_G$ | $\Gamma_D$ (cm$^{-1}$) | $\Gamma_G$ (cm$^{-1}$) | $\Gamma_{2D}$ (cm$^{-1}$) |
|---|---|---|---|---|
| 20 | 2.6±0.3 | 42±5 | 43±6 | 74±12 |
| 30 | 2.8±0.3 | 39±4 | 41±4 | 67±10 |
| 40 | 2.7±0.3 | 41±5 | 42±5 | 70±12 |
| 50 | 2.8±0.3 | 37±4 | 40±5 | 64±9 |
| 75 | 2.6±0.3 | 43±6 | 42±6 | 75±13 |
| 100 | 2.8±0.3 | 40±5 | 41±5 | 72±13 |
| 150 | 1.4±0.2 | 79±12 | 80±54 | 182±66 |
| 200 | 1.2±0.1 | 117±9 | 74±4 | 225±26 |
| 500 | 1.1±0.1 | 167±10 | 84±4 | 301±40 |
| 1000 | 1.4±0.1 | 66±7 | 146±72 | 127±40 |

[1]   S. Eigler, C. Dotzer, A. Hirsch, M. Enzelberger, P. Müller, *Chem. Mater.* **2012**, *24*, 1276-1282.